\documentclass[conference]{IEEEtran}
\IEEEoverridecommandlockouts
\usepackage{cite}
\usepackage{amsmath,amssymb,amsfonts}
\usepackage{algorithmic}
\usepackage[ruled,vlined]{algorithm2e}
\usepackage{multirow}
\usepackage{graphicx}
\usepackage{textcomp}
\usepackage{xcolor}
\def\BibTeX{{\rm B\kern-.05em{\sc i\kern-.025em b}\kern-.08em
    T\kern-.1667em\lower.7ex\hbox{E}\kern-.125emX}}
\begin{document}

\title{Intra-Day Price Simulation with Generative Adversarial Modelling of the Order Flow}

\author{\IEEEauthorblockN{Ye-Sheen Lim}
\IEEEauthorblockA{\textit{Department of Computer Science} \\
\textit{University College London}\\
London, United Kingdom \\
y.lim@cs.ucl.ac.uk}
\and
\IEEEauthorblockN{Denise Gorse}
\IEEEauthorblockA{\textit{Department of Computer Science} \\
\textit{University College London}\\
London, United Kingdom \\
d.gorse@cs.ucl.ac.uk}
}

\maketitle

\begin{abstract}
Intra-day price variations in financial markets are driven by the sequence of orders, called the order flow, that is submitted at high frequency by traders. This paper introduces a novel application of the Sequence Generative Adversarial Networks framework to model the order flow, such that random sequences of the order flow can then be generated to simulate the intra-day variation of prices. As a benchmark, a well-known parametric model from the quantitative finance literature is selected. The models are fitted, and then multiple random paths of the order flow sequences are sampled from each model. Model performances are then evaluated by using the generated sequences to simulate price variations, and we compare the empirical regularities between the price variations produced by the generated and real sequences. The empirical regularities considered include the distribution of the price log-returns, the price volatility, and the heavy-tail of the log-returns distributions. The results show that the order sequences from the generative model are better able to reproduce the statistical behaviour of real price variations than the sequences from the benchmark.
\end{abstract}

\section{Introduction}

Most of today's fast-paced markets are organised as \emph{order-driven} markets, examples being found among the world's largest equity exchanges, such as the NASDAQ, the NYSE, Hong Kong, Shanghai, Shenzhen, London and Toronto Stock Exchange, and the Euronext. The \emph{order flow} is the dynamic sequence of orders submitted by traders in an order-driven market, and is the lever that causes the variation of prices at high-frequency timescales, such as intra-day price variations.

The state-of-the-art in the application of deep learning for modelling high-frequency price variations has focused on directly predicting the directional price change \cite{tsantekidis2017using,tsantekidis2017forecasting,dixon2018sequence,passalis2018temporal,dixon2018sequence,sirignano2018universal,lim2020deep}. Although this previous work has reported promising results, there are a number of advantages in addressing the generative modelling of the order flow itself, including the computation of future order intensities for high-frequency trading strategies \cite{avellaneda2008high}, the provision of data-driven insights into the market microstructure \cite{o2015high}, and as a simulator for evaluating and back-testing trading strategies \cite{hu2014agent}, with the advantage addressed here being the simulation of future price variations using the generated order sequences.

To our knowledge there is currently a gap in the machine learning literature in applying deep learning, or any machine learning models, to modelling the order flow. This paper fills this gap by introducing the \emph{Sequence Generative Adversarial Network} (SeqGAN) \cite{yu2017seqgan} for modelling the order flow. Since there is currently no related work in the machine learning literature, a well-known model from the quantitative finance literature is selected as benchmark. Model performance is evaluated by performing a statistical analysis of the simulated intra-day price variation resulting from the generated sequences, and comparing the results to ones acquired from corresponding analyses of real data.

\section{Related Work}
\label{section:related_work}

There are two philosophies of statistical modelling when deriving conclusions from data. One assumes a data generating process, while the other uses algorithmic models that treat the data mechanism as unknown. In modelling order flow data, the former approach gives rise to the \emph{zero-intelligence} models in the quantitative finance literature, while machine learning models fall into the latter class of algorithmic models.

Zero-intelligence models assume that the order flow is governed by stochastic processes without any assumptions about rational trader behaviour. The current state-of-the-art uses a framework that models the irregularly-spaced market, limit, and cancellation orders using independent counting processes. Most commonly the multiple Poisson process is used, where each process models the independent arrival of an order at a given price \cite{smith2003statistical,cont2010stochastic}. These zero-intelligence models are able to reproduce many empirical regularities found in real price variations. However, due to simplified assumptions about the data-generating mechanism, these models are sensitive to regime shifts, and lack generalisation power. Tractability and parametric estimation can also be an issue.

These drawbacks may be overcome by machine learning approaches that learn directly from data without assuming any data-generating process. However, to our knowledge, no previous work has applied machine learning to the modelling of the order flow. The closest related work that applies deep learning to order flow data is in \cite{tsantekidis2017using,tsantekidis2017forecasting,dixon2018sequence,passalis2018temporal,dixon2018sequence,sirignano2018universal,lim2020deep}, which predict high-frequency price variations using the limit order book or order flow related data, though they do not address the problem of modelling the order flow sequences. \cite{zhang2019stock,zhou2018stock,takahashi2019modeling} have applied GANs to directly model price time-series; however, in this paper, we are instead interested in learning the data-generating process that produces these price time-series.

\section{Domain Background}
\label{section:domain_background}

In this section, some features of order-driven markets are briefly introduced. Readers are directed to \cite{gould2013limit} for further technical details of order-driven markets.

The order flow is the sequence of placement and cancellation of \emph{limit orders} and \emph{market orders} by traders in order-driven markets. A limit order is a type of order to buy or sell a volume of a traded asset at a specified price or better. Although the price is guaranteed, the filling of a limit order is not. If a submitted limit order cannot be immediately executed against an existing order in the \emph{limit order book} (LOB), then the limit order is added to the LOB until it is cancelled, amended, or executed against subsequent orders. Meanwhile, market orders are immediately executed against limit orders queued at the best price in the order book, as fully as possible. Any unfilled portion may then be converted to limit orders at the same price, or executed at the next best available price until the market order is fully executed.

\begin{figure}[ht!]
    \centering
    \caption{A sample visualisation of a limit order book.}
    \includegraphics[width=0.99\columnwidth]{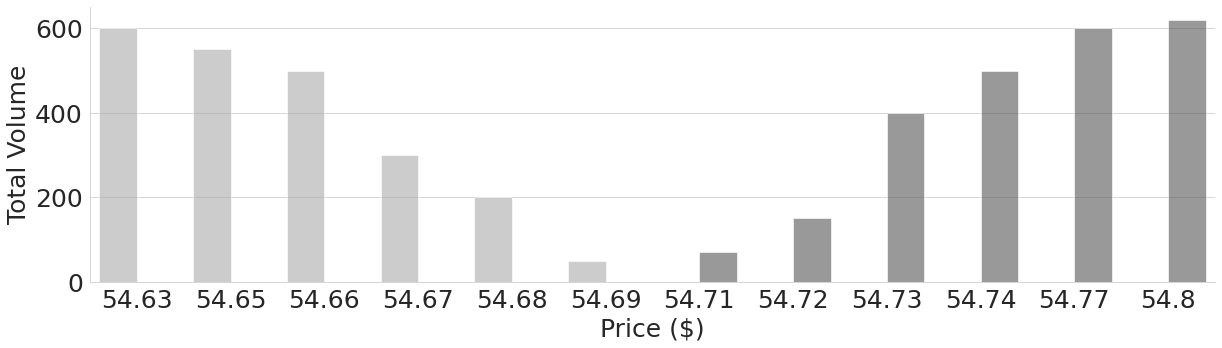}
\label{section:domain_background:fig:lob}
\end{figure}

At any given time, the total volumes of limit orders in the LOB are grouped by price. This is the common state of the LOB as visualised and evaluated by traders. An example is shown in Figure \ref{section:domain_background:fig:lob}. Buy limit orders are on the left side, sorted by the highest price to the right, while sell limit orders are on the right side, sorted by the lowest price to the left.

Some necessary common terminology for various measures in the LOB must now be defined. The highest buy (or, lowest sell) price in the LOB at time $t$ is the \emph{best bid} $b(t)$ (or, \emph{best ask} $a(t)$). The \emph{mid-price} is $\frac{a(t)+b(t)}{2}$, and is the most common reference price when trading at high-frequency time-scales. In the example of Figure \ref{section:domain_background:fig:lob}, the mid-price is \$54.70. Finally, we will need to define the \emph{relative price}. Since the prices in an LOB are constantly changing, it is more useful to have a relative measure of the price rather than any specific price. From a modelling perspective, this will also naturally normalise any price variables. The relative price a buy (or, sell) order at time $t$ is the number of \emph{ticks} from $a(t)$ (or, $b(t)$), where a tick is the smallest permissible price change imposed by the exchange. In the example of Figure \ref{section:domain_background:fig:lob}, and assuming a tick of \$0.01, the relative price for the orders at \$54.68 is exactly 3 since it is 3 ticks away from $a(t)=\$54.71$. 

\section{Technical Background}
\label{section:technical_background}

Though recurrent neural networks (RNNs) have been successful in modelling sequence data \cite{graves2013generating}, RNNs trained with maximum likelihood suffer from \emph{exposure bias} in the inference stage when generating sequences \cite{bengio2015scheduled}. The generative adversarial networks (GAN) framework \cite{goodfellow2014generative} is a potential solution to the exposure bias problem. GANs are made up of a generator $G_\theta$ and discriminator $D_\phi$, parameterised by $\theta$ and $\phi$ respectively. The generator $G_\theta$ is trained to produce a sequence $Y_{1:T} = (y_1, \dots, y_T )$, where $y_t \in \mathcal{Y}$ and $\mathcal{Y}$ is some set of discrete tokens. The discriminator $D_\phi$ is a binary classifier trained to distinguish the generated sequence $Y_{1:T}$ from a real sequence $X_{1:T}$. The probability $D_\phi(Y_{1:T})$, which measures how likely it is that generated sequence is a real sequence, is used as feedback for guidance as to how to further improve $G_\theta$. However, standard GANs are designed for real-valued continuous data, which is not appropriate for the order flow data considered in this paper, since the orders are naturally discrete event tokens.

The work in \cite{yu2017seqgan} addresses the adversarial learning of discrete sequences by introducing the SeqGAN framework. In a SeqGAN, the training of the generator is treated as a reinforcement learning (RL) problem. At any given timestep $t$, the state $s$ is the sequence produced thus far, $y_1,\dots,y_{t-1}$. The action $a$ is then which token to select as $y_t$ from $\mathcal{Y}$. The action to be taken in a given state is determined by the generator $G_\theta(y_t|Y_{1:t-1})$, which is a stochastic policy parameterised by $\theta$. The generator is updated via policy gradient, utilising rewards in the form of the output of the discriminator $D_\phi$. The action value defined by the authors of \cite{yu2017seqgan} for updating the generator is as follows:

\begin{equation}
  \begin{aligned}
    Q^{G_\theta}_{D_\phi}&(s=Y_{1:t-1},a=y_t) = \\
    &\begin{cases}
      \frac{1}{N}\sum^N_{n=1} D_\phi(Y^n_{1:T}) & \text{if } t<T, \\
      D_\phi(Y_{1:t}),              & \text{if } t=T,
    \end{cases}
  \end{aligned}
  \label{section:technical_background:eq:action_value}
\end{equation}

\noindent
where $Y^n_{1:T}\in\text{MC}^{G_\theta}(Y_{1:t};N)$, $\text{MC}^{G_\beta}(Y_{1:t};N)$ represents an $N$-times Monte Carlo search algorithm for sampling the unknown last $T-t$ tokens using the generator $G_\theta$ as the rollout policy, and $Y^n_{1:T}$ is a sampled sequence. The gradient for the generator objective $J(\theta)$ is computed as:

\begin{equation}
  \nabla_\theta J(\theta) \approx \sum^T_{t=1} \nabla_\theta \log G_\theta(y_t|Y_{1:t-1}) \cdot Q^{G_\theta}_{D_\phi}(Y_{1:t-1,y_t}),
  \label{section:technical_background:eq:dj_dt}
\end{equation}

After improving the generator via policy gradient update for a number of epochs, $D_\phi$ is re-trained using negative examples produced from the improved $G_\theta$ by minimising the binary cross-entropy loss. Readers are directed to the original paper \cite{yu2017seqgan} for more detailed explanations and derivations.

\section{Methodology}
\label{section:methodology}

\subsection{Problem Formulation}
\label{section:methodology:subsection:problem_formulation}

An order in the order flow is defined in this paper to be a \emph{discrete event token} that represents a buy or sell market, limit, or cancellation order at a given relative price $q$. Let $\mathcal{O}$ be this set of event tokens, defined as follows:

\begin{gather}
    \mathcal{O} \in \{ \mathcal{L} \cup \mathcal{M} \cup \mathcal{C} \cup \mathcal{E} \}, \\
    \mathcal{L} \in \{ l_{B,1}, \dots, l_{B,Q},l_{A,1}, \dots, l_{A,Q} \}, \\
    \mathcal{C} \in \{ c_{B,1}, \dots, c_{B,Q}, c_{A,1}, \dots, c_{A,Q} \}, \\
    \mathcal{M} \in \{ \mu_{B}, \mu_{A} \}, \\
    \mathcal{E} \in \{ \eta_B, \eta_A \},
\end{gather}

\noindent
where $\mathcal{L}$ is the set of all limit order event tokens, $\mathcal{M}$ is the set of all market order event tokens, $\mathcal{C}$ is the set of all cancellation event tokens, and $\mathcal{E}$ is the set of all other event tokens.

Each event token in $\mathcal{L}$, $\mathcal{M}$, $\mathcal{C}$, and $\mathcal{E}$, can be described as follows. The token $l_{B,q}$ represents a bid limit order at a relative price $q$ from the best ask, while $l_{A,q}$ represents an ask limit order at a relative price $q$ from the best bid. Using a similar notation, $c_{B,q}$ and $c_{A,q}$ are tokens for the cancellation of active bid and ask limit orders. All limit and cancellation orders not within $Q$ relative prices are represented by a single token $\eta_A$ for the ask side, and $\eta_B$ for the bid side of the order book, respectively. Finally, market orders arriving at the best bid and best ask are represented by $\mu_{B}$ and $\mu_{A}$ respectively. For the limit and cancellation orders, prices more than $Q$ ticks away from the best bid and best ask are not considered, since trading activities that impact the market occur mostly at prices closer to the best bid and best ask \cite{cont2014price}.

Given this set $\mathcal{O}$ of order event tokens, the order flow is defined here as a finite-length sequence $O_{1:T} = \{o_1, \dots, o_T \}$, where $o_{t} \in \mathcal{O}$ is a token indicating the type of order event arriving at a given relative price. Therefore we have here a discrete token sequence modelling problem. We aim to train $G_\theta$ on $O_{1:T}$ to produce a novel sequence of orders $O'_{1:T} = \{o'_1,\dots,o'_T\}$ such that the probabilistic difference between generated and real sequences is minimised.

\subsection{SeqGAN Modelling of Order Flow Sequences}

The algorithm for training the generator $G_\theta$ and discriminator $D_\phi$ using the SeqGAN framework is described in Algorithm \ref{section:methodology:subsection:seqgan_order:algorithm:token_train}. A recurrent neural network (RNN) with long-short term memory cells \cite{hochreiter1997long} is implemented as the generator model $G_\theta$, while the discriminator $D_\theta$ is implemented by a convolutional neural network \cite{zhang2015sensitivity}. In both networks, the tokens $(y_1,\dots,y_T)$ are embedded into a continuous space $(x_1,\dots,x_T)$ using a fully connected layer.

\begin{algorithm}[ht!]
\LinesNumbered
  \KwIn{\\
    \Indp
    Order flow real sequences $X = \{O_{1:T}\}_{1:N}$; \\ 
    Order flow start sequences $S = \{O_{-T+1:0}\}_{1:N}$
  }
  Initialise $G_\theta$ and $D_\phi$ with random parameters $\theta$ and $\phi$\;
  Pre-train $G_\theta$ using MLE on $X$ with starting sequences $S$\;
  Generate samples using $G_\theta$ using starting state $S$\;
  Pre-train $D_\theta$ by minimising CE on generated samples as negative examples and $X$ as positive examples\;
  \Repeat(){SeqGAN converges}{
    \For{g-steps}{
      Uniformly sample starting sequence $s$ from $S$\;
      Generate ${o'_1,\dots,o'_T}$ using $G_\theta$ with starting state $s$\;
      \For{$t$ in $1:T$}{
        Compute $Q(a=o'_t, s=O'_{1:t-1})$ using Eq. \ref{section:technical_background:eq:action_value}\;
      }
      Update $\theta$ using Eq. \ref{section:technical_background:eq:dj_dt}\;
    }
    \For{d-steps}{
      \For{each $s$ in $S$}{
        Generate sequence sample using $G_\theta$ with starting state $s$\;
        Append sequence sample to array of negative examples\;
      }
      Uniformly sample equal number of negative examples and positive examples $X$\;
      Use bootstrapped data to train $D_\theta$ for a number of epochs given by Eq. \ref{section:technical_background:eq:action_value}\;
    }
  }
\caption{Algorithm for training the SeqGAN generator and discriminator on the order flow.}
\label{section:methodology:subsection:seqgan_order:algorithm:token_train}
\end{algorithm}

In the original SeqGAN paper \cite{yu2017seqgan}, the start state $s_0$ in the SeqGAN framework is a special token defining the start of a sequence, commonly used in natural language processing datasets. For the work here, it is proposed that the start state be a sequence of order flow, our reason being that it would be unnatural for an order flow to abruptly start, unlike a text sentence. Therefore, to generate an order flow $O'_{1:T}$, a start sequence is defined as $O_{-T+1:0}=\{o_{-T+1,-T+2,\dots,o_0} \}$, where the length of the start sequence is set to be the same as the length of the sequence to be generated. A given start sequence is always associated with a positive sequence example in the training set such that $O_{-T+1:0}$ concatenated with a positive example $O_{1:T}$ forms a continuous sequence of order flow that exists in the real data. When generating a simulated sequence $O'_{1:T}$, the start state can be uniformly sampled with replacement from the set of start sequences.

\subsection{Benchmark Model}
\label{section:methodology:subsection:benchmark_model}

Since the generative modelling of order flow sequences in this paper is novel, no direct comparison with existing approaches in the machine learning literature can be made. However, the quantitative finance literature contains well-known stochastic process approaches for the modelling of order flow, as presented in Section \ref{section:related_work}. Among these, the \emph{multiple Poisson process} \cite{smith2003statistical,cont2010stochastic} is the most suitable and reliable to use here as a benchmark model, due to its ubiquity in practice, as well as its simplicity in parameter estimation. In the multiple Poisson model, one Poisson process is used for modelling the arrival of a single order event tokenised in set $\mathcal{O}$. Interested readers are directed to \cite{cont2010stochastic} for more details of the model and how it is fitted. After the arrival rate parameter for each of the processes is fitted, the procedure for generating a sequence of tokens is quite straightforward. For each process, the arrival time of the token is sampled from the process. Then, all of the generated token sequences are concatenated into a single data-structure and sorted by time to obtain the generated order flow.

\section{Dataset}
\label{section:dataset}

Order flow data from most stock exchanges is either very expensive or difficult to obtain for the typical researcher. However, cryptocurrency exchanges allow access to the same kind of order flow data as could be obtained from regular stock exchanges, but at virtually no cost. The data for our experiments are for this reason obtained from Coinbase, a digital currency exchange. In this paper we gathered the order flow for the BTC-USD currency pair in the period between 4 Nov 2017 to 1 Dec 2017.

Data from the period between 4 Nov and 29 Nov is used for training. In this period, the order flow is partitioned into slices of 400 events. Each of the slices is split equally into two to obtain the real order flow sequence $\hat{O}_{1:T}$ and the start sequence $\hat{O}_{-T+1:0}$. All of the real order flow sequences are concatenated into a single dataset for training the discriminator $D_\phi$, while the start sequences are concatenated into a dataset to be used for generating a sequence of order flow in the generator $G_\theta$. For testing, we use the data from the period between 30 Nov and 1 Dec. 

\section{Results: Macro-Behaviour Analyses}
\label{section:results:subsection:macro-behaviour}

For model evaluation, a set of macro-behaviour analyses are conducted to investigate how well the intra-day mid-price variations of the simulated order flow from both models reproduce important empirical regularities found in real mid-price variations. Specifically, this section will compare the mid-price log-returns distribution and the mid-price volatility for both models to that of the real mid-price series, over the test period between 30 Nov and 1 Dec 2017.

To simulate the intra-day price variation, the order volume and inter-order arrival time for each generated order needs to be sampled from an empirical distribution. The benchmark model needs only to sample the order volume, since the multiple Poisson model naturally models the arrival time of each order. The empirical distributions for the order volume and inter-order arrival time are estimated from the data in the training period. We generate enough order flow data from each model to produce a 1 minute interval mid-price time series for a period of 48 hours.

\subsection{Mid-Price Returns Distribution}

We first compare the log-returns distributions from the simulated mid-prices to that for the real mid-price time series using the two-sample Kolmogorov-Smirnov (K-S) test. Denoting the dataset of log-returns computed from one sample of a simulated time series as $A$, and the dataset of log-returns from the real mid-price series as $B$, the K-S test is then performed under the null hypothesis that datasets $A$ and $B$ are sampled from the same distribution. Since 100 samples of the simulated order flow sequences were obtained for both models, the K-S test has to be performed 100 times for each model.

However, we now encounter the issue of multiple comparison since the more samples of the simulated mid-prices we test, the more likely it is that one of them would pass the K-S test. To avoid this bias, Hochberg's step-up procedure \cite{dunnett1992step} is implemented as an additional step to control the outcome of the multiple K-S tests. The procedure sorts the hypotheses of the 100 K-S tests by p-value, and determines which of the hypotheses, those with the lowest p-values, should be rejected. For these tests, a larger than usual significance level of 0.1 is chosen since simulating noisy financial time-series is an immense challenge. Then, comparing the SeqGAN model and the benchmark, we say that the model with the least number of hypotheses rejected by Hochberg's step-up procedure is that which is more likely to produce an order flow with realistic macro-behaviour.

\begin{table}[ht!]
\centering
\caption{Number of Kolmogorov-Smirnov test hypotheses (out of 100 samples) rejected in Hochberg's step-up procedure. The length column refers to the first 1, 6 and 48 hours for each of the 100 samples.}
\begin{tabular}{||l|c|c||}
\hline
Time-Series Length & SeqGAN Model & Benchmark Model \\ \hline \hline
1 Hour     & \textbf{73}            & 86              \\
6 Hours    & \textbf{88}            & 91              \\
48 Hours   & \textbf{98}            & 100             \\ \hline
\end{tabular}
\label{section:results:subsection:macro-behaviour:table:hypo_reject}
\end{table}

Table \ref{section:results:subsection:macro-behaviour:table:hypo_reject} shows the number of hypotheses rejected for the SeqGAN model and benchmark model, with the experiments replicated for the first 1 hour, 6 hours, and 48 hours of the mid-price time-series. It can be observed that as the time-series length is increased, the similarity between the log-return distributions of the simulated order flow and the real order flow deteriorates, as would be expected. For the longer time-series, quite a large number of the samples are rejected for both models, but this is again as expected since high-frequency financial time-series are extremely challenging to realistically replicate, especially for long time periods.

Recall that the simulated order flow for the mid-price time series is produced iteratively where, initially, a new simulated sequence is generated from a starting sequence of real order flow. This generated sequence is then fed back as a starting sequence to generate another new sequence, and so on. The performance for time-series of different lengths in Table \ref{section:results:subsection:macro-behaviour:table:hypo_reject} would suggest that as each new sequence is generated, conditioned on a previously generated sequence, the resulting statistical behaviour of the mid-price log-returns starts to deviate from that of the actual ones. Although Generative Adversarial Networks in theory mitigate this exposure bias problem, it seems as if for this experiment the problem even so persists in the long run.

Nonetheless, the results here show the simulated order flow produced by the SeqGAN model is better at reproducing the mid-price log-returns of real data than the benchmark for all three time-series lengths in the experiment.

\subsection{Mid-Price Tail Exponents}

Next, the tails of the absolute log-returns distributions for the simulated mid-price of each of the models are compared to those of the real data. Empirical studies have reported strong evidence of power law behaviour \cite{gould2013limit} in the absolute log-return distributions of financial time series. Power law probability distributions are “heavy-tailed”, meaning the right tails of the distributions still contain a great deal of probability. Power law distributions are probability distributions with the form $p(x) \propto x^{-\alpha}$, and it is the tail-exponent $\alpha$ that is the subject of this analysis in this section.

The Jarque-Bera (JB) test \cite{jarque1980efficient} is first applied to the real mid-price series, for the first 1 hour, 6 hours and 48 hours, to determine if there are heavy tails in the absolute log-returns distribution. From Table \ref{section:results:subsection:macro-behaviour:table:fat_tail}, it can be observed that the kurtosis of the test distributions is much larger than 3, indicating heavy tails with very high statistical significance. 

\begin{table}[ht!]
\centering
\caption{Test period kurtosis and p-values from the Jarque-Bera test, and computed tail-exponents, for the real mid-price time-series absolute log-returns. The length column refers to the first 1, 6, and 48 hours of the series.}
\begin{tabular}{||l|ccc||}
\hline
Time-Series Length & Tail-Exponent & Kurtosis & p-value \\ \hline \hline
1 Hour     & 3.67             & 8.79     & 0.00                \\
6 Hour     & 2.98             & 8.46     & 0.00                \\
48 Hour    & 3.30             & 10.98    & 0.00                \\ \hline
\end{tabular}
\label{section:results:subsection:macro-behaviour:table:fat_tail}
\end{table}

We then equivalently test the absolute log-returns of the simulated mid-price series for heavy tails. Table \ref{section:results:subsection:macro-behaviour:table:fat_tail_models} shows the aggregated results of the Jarque-Bera test across the 100 samples generated by the SeqGAN and benchmark models. The measured kurtoses are averaged, and the Hochberg Step-Up procedure is again applied to determine the proportion of tests to be accepted at a 1\% significance, after controlling for repetition bias. It can be observed that the average kurtosis is much larger than 3, and the null hypotheses of the Jarque-Bera test across all samples are rejected. The results thus strongly indicate that the simulated mid-price series for both the SeqGAN and the benchmark models do replicate the heavy tails reported for real financial time-series.

\begin{table}[ht!]
\centering
\caption{Mean kurtosis from the Jarque-Bera test, and the number of tests rejected by the Hochberg Step-Up procedure, for the 100 mid-price time-series samples generated by the SeqGAN and benchmark models. The length column refers to the first 1, 6, and 48 hours for each of the 100 samples.}
\begin{tabular}{||l|l|cc||}
\hline
Length        & Metric          & SeqGAN Model & Benchmark Model \\ \hline \hline
\multirow{2}{*}{1 Hour}   & Mean Kurtosis   & 7.31           & 6.97            \\
                          & Rejection Count & 0              & 0               \\ \hline
\multirow{2}{*}{6 Hours}  & Mean Kurtosis   & 7.49           & 7.19            \\
                          & Rejection Count & 0              & 0               \\ \hline
\multirow{2}{*}{48 Hours} & Mean Kurtosis   & 8.80           & 7.53            \\
                          & Rejection Count & 0              & 0               \\ \hline
\end{tabular}
\label{section:results:subsection:macro-behaviour:table:fat_tail_models}
\end{table}

We next compare the tail-exponents of the simulated mid-price distribution to those of the real data in Table \ref{section:results:subsection:macro-behaviour:table:fat_tail}. First, the distribution of tail-exponents is computed for the sampled mid-price time-series of each model. We then apply the one-sample two-tailed Student t-test between the tail-exponents from the simulated mid-price series and real mid-price series. Table \ref{section:results:subsection:macro-behaviour:table:tail_expo} shows the resulting p-value and t-statistics. We see that the null hypotheses of the tests for both models are rejected with high confidence, implying that neither model was realistically producing the tail-exponents. It can also be observed from the t-statistics that the distribution tails of both models are lighter than those of the real returns distribution. However, the t-statistics of the SeqGAN model are much smaller that those of the benchmark model, indicating that the SeqGAN model simulates mid-price variations with closer tail-exponent behaviours to those of the real data.

\begin{table}[ht!]
\centering
\caption{Test period results of one-sample two-tailed Student t-tests for the tail-exponent distributions of each model against the real tail-exponents, rounded to two decimal places. The length column refers to the first 1, 6 and 48 hours for each of the 100 samples.}
\begin{tabular}{||l|l|c|c||}
\hline
Length        & Metric      & SeqGAN Model & Benchmark Model \\ \hline \hline
\multirow{2}{*}{1 Hour}   & p-value     & 0.00           & 0.00            \\
                          & t-statistic & \textbf{2.66}  & 3.29            \\ \hline
\multirow{2}{*}{6 Hours}  & p-value     & 0.00           & 0.00            \\
                          & t-statistic & \textbf{2.84}  & 4.05            \\ \hline
\multirow{2}{*}{48 Hours} & p-value     & 0.00           & 0.00            \\
                          & t-statistic & \textbf{2.23}  & 3.71            \\ \hline
\end{tabular}
\label{section:results:subsection:macro-behaviour:table:tail_expo}
\end{table}

\subsection{Mid-Price Volatility}

Finally, the volatility in the mid-price produced by the two models is compared to the volatility of the real mid-price in the test set. Volatility is one of the most important measures of an asset's value as it measures the risk that would be undertaken when trading the security and is crucial in the construction of optimal portfolios. There are a number of measures of volatility defined in the literature, and the choice depends on the purpose. For intra-day price variations, the volatility definitions that have significance importance are the \emph{realised volatility} $v_r$, the \emph{realised volatility per trade} $v_p$, and the \emph{intraday volatility} $v_d$, as described in more detail in \cite{gould2013limit}.

Table \ref{section:results:subsection:macro-behaviour:table:real_vol} shows each of these volatilities computed from the real mid-price in the test period. A comparison to the mid-price volatilities of the SeqGAN and benchmark models is performed as follows. First, the empirical distributions of the volatility measures are computed across the simulated mid-price series produced by each model. A one-sample two-tailed Student t-test is then applied between the the data in the empirical distributions, and the volatility computed from the real mid-price. The results from the tests are given in Table \ref{section:results:subsection:macro-behaviour:table:compare_vol}, where it can be observed that the null hypotheses for all the tests are rejected with high confidence. This implies that neither model can generate an order flow able to reproduce the volatility of the real mid-price time series, with the negative t-statistics implying that the simulated mid-price time-series have much lower volatility than the real time-series. However, comparing the SeqGAN and benchmark models, it can be observed from the t-statistics in Table \ref{section:results:subsection:macro-behaviour:table:compare_vol} that the volatility of the mid-price is in general better replicated by the SeqGAN model. An exception to this would be the intraday volatility for time-series lengths 6 hours and 48 hours, which the t-statistics show were reproduced more closely by the benchmark model than the SeqGAN model. 

\begin{table}[ht!]
\centering
\caption{Volatility measures computed from the real mid-price time-series in the test period. The length column refers to the first 1, 6, and 48 hours for each of the 100 samples.}
\begin{tabular}{||l|c|c|c||}
\hline
Time-Series Length & $v_r$ & $v_p$ & $v_d$ \\ \hline \hline
1 Hour             & 0.00177             & 0.00149                       & 0.0308              \\
6 Hours            & 0.00186             & 0.00153                       & 0.099               \\
48 Hours           & 0.00257             & 0.00211                       & 0.178  \\    \hline       
\end{tabular}
\label{section:results:subsection:macro-behaviour:table:real_vol}
\end{table}

\begin{table}[ht!]
\centering
\caption{p-value and t-statistics of the one-sample two tailed Student t-test between different volatility distributions of the SeqGAN and benchmark models, against the real volatility measures, rounded to two decimal places. The length column refers to the first 1, 6, and 48 hours for each of the 100 samples.}
\begin{tabular}{||l|l|cc|cc||}
\hline
\multirow{2}{*}{Length} & \multirow{2}{*}{Volatility} & \multicolumn{2}{c||}{SeqGAN Model} & \multicolumn{2}{c|}{Benchmark Model} \\ \cline{3-6} 
                                    &                                     & t-statistics         & p-value      & t-statistics         & p-value       \\ \hline \hline
\multirow{3}{*}{1 Hour}             & $v_r$                               & \textbf{-0.92}       & 0.00         & -0.99                & 0.00          \\
                                    & $v_p$                               & \textbf{-0.99}       & 0.00         & -1.10                & 0.00          \\
                                    & $v_d$                               & \textbf{-0.89}       & 0.00         & -0.93                & 0.00          \\ \hline
\multirow{3}{*}{6 Hours}            & $v_r$                               & \textbf{-1.04}       & 0.00         & -1.13                & 0.00          \\
                                    & $v_p$                               & \textbf{-0.99}       & 0.00         & -1.19                & 0.00          \\
                                    & $v_d$                               & -1.11                & 0.00         & \textbf{-0.95}       & 0.00          \\ \hline
\multirow{3}{*}{48 Hours}           & $v_r$                               & \textbf{-1.32}       & 0.00         & -1.46                & 0.00          \\
                                    & $v_p$                               & \textbf{-1.27}       & 0.00         & -1.43                & 0.00          \\
                                    & $v_d$                               & -1.18                & 0.00         & \textbf{-1.03}       & 0.00          \\ \hline
\end{tabular}
\label{section:results:subsection:macro-behaviour:table:compare_vol}
\end{table}

\section{Conclusion}
\label{section:conclusion}

A novel application of the SeqGAN framework for generating simulated order flow sequences was introduced and benchmarked against a well-known model from the quantitative finance literature. An analysis of the macro-behaviour of the mid-price movements showed that the SeqGAN model is substantially better able than the benchmark model to replicate the overall returns distribution, the returns distribution tails and the volatility of the real mid-price time-series. While the results showed that there is further work to be done to improve this approach to the generative modelling of the order flow, financial sequences are in general hard to predict and even harder to simulate. Future work that could improve the generative modelling of order sequences could include improving the architecture to inject other covariates into the input of the generator, or extending the architecture to jointly model the event tokens, order volume, and inter-order arrival times. Also, further analysis on the actual sequences that were generated by the models could determine what is needed to improve the model. On the basis of its current performance, and with further work along these lines, including comparison of this method against an increased number of benchmarks, we believe the SeqGAN model could be of substantial practical value to the financial community in the future.

\bibliographystyle{IEEEtran}
\bibliography{ref}

\begin{thebibliography}{10}
\providecommand{\url}[1]{#1}
\csname url@samestyle\endcsname
\providecommand{\newblock}{\relax}
\providecommand{\bibinfo}[2]{#2}
\providecommand{\BIBentrySTDinterwordspacing}{\spaceskip=0pt\relax}
\providecommand{\BIBentryALTinterwordstretchfactor}{4}
\providecommand{\BIBentryALTinterwordspacing}{\spaceskip=\fontdimen2\font plus
\BIBentryALTinterwordstretchfactor\fontdimen3\font minus
  \fontdimen4\font\relax}
\providecommand{\BIBforeignlanguage}[2]{{%
\expandafter\ifx\csname l@#1\endcsname\relax
\typeout{** WARNING: IEEEtran.bst: No hyphenation pattern has been}%
\typeout{** loaded for the language `#1'. Using the pattern for}%
\typeout{** the default language instead.}%
\else
\language=\csname l@#1\endcsname
\fi
#2}}
\providecommand{\BIBdecl}{\relax}
\BIBdecl

\bibitem{tsantekidis2017using}
A.~Tsantekidis, N.~Passalis, A.~Tefas, J.~Kanniainen, M.~Gabbouj, and
  A.~Iosifidis, ``Using deep learning to detect price change indications in
  financial markets,'' in \emph{Signal Processing Conference (EUSIPCO), 2017
  25th European}.\hskip 1em plus 0.5em minus 0.4em\relax IEEE, 2017, pp.
  2511--2515.

\bibitem{tsantekidis2017forecasting}
------, ``Forecasting stock prices from the limit order book using
  convolutional neural networks,'' in \emph{2017 IEEE 19th Conference on
  Business Informatics (CBI)}, vol.~1.\hskip 1em plus 0.5em minus 0.4em\relax
  IEEE, 2017, pp. 7--12.

\bibitem{dixon2018sequence}
M.~Dixon, ``Sequence classification of the limit order book using recurrent
  neural networks,'' \emph{Journal of computational science}, vol.~24, pp.
  277--286, 2018.

\bibitem{passalis2018temporal}
N.~Passalis, A.~Tefas, J.~Kanniainen, M.~Gabbouj, and A.~Iosifidis, ``Temporal
  bag-of-features learning for predicting mid price movements using high
  frequency limit order book data,'' \emph{IEEE Transactions on Emerging Topics
  in Computational Intelligence}, 2018.

\bibitem{sirignano2018universal}
J.~Sirignano and R.~Cont, ``Universal features of price formation in financial
  markets: perspectives from deep learning,'' \emph{Quantitative Finance},
  vol.~19, no.~9, pp. 1449--1459, 2019.

\bibitem{lim2020deep}
Y.-S. Lim and D.~Gorse, ``Deep probabilistic modelling of price movements for
  high-frequency trading,'' in \emph{2020 International Joint Conference on
  Neural Networks (IJCNN)}.\hskip 1em plus 0.5em minus 0.4em\relax IEEE, 2020,
  pp. 1--8.

\bibitem{avellaneda2008high}
M.~Avellaneda and S.~Stoikov, ``High-frequency trading in a limit order book,''
  \emph{Quantitative Finance}, vol.~8, no.~3, pp. 217--224, 2008.

\bibitem{o2015high}
M.~O’Hara, ``High frequency market microstructure,'' \emph{Journal of
  Financial Economics}, vol. 116, no.~2, pp. 257--270, 2015.

\bibitem{hu2014agent}
R.~Hu and S.~M. Watt, ``An agent-based financial market simulator for
  evaluation of algorithmic trading strategies,'' in \emph{6th International
  Conference on Advances in System Simulation}.\hskip 1em plus 0.5em minus
  0.4em\relax Citeseer, 2014, pp. 221--227.

\bibitem{yu2017seqgan}
L.~Yu, W.~Zhang, J.~Wang, and Y.~Yu, ``Seqgan: Sequence generative adversarial
  nets with policy gradient,'' in \emph{Proceedings of the AAAI conference on
  artificial intelligence}, vol.~31, 2017.

\bibitem{smith2003statistical}
E.~Smith, J.~D. Farmer, L.~s. Gillemot, S.~Krishnamurthy \emph{et~al.},
  ``Statistical theory of the continuous double auction,'' \emph{Quantitative
  finance}, vol.~3, no.~6, pp. 481--514, 2003.

\bibitem{cont2010stochastic}
R.~Cont, S.~Stoikov, and R.~Talreja, ``A stochastic model for order book
  dynamics,'' \emph{Operations research}, vol.~58, no.~3, pp. 549--563, 2010.

\bibitem{zhang2019stock}
K.~Zhang, G.~Zhong, J.~Dong, S.~Wang, and Y.~Wang, ``Stock market prediction
  based on generative adversarial network,'' \emph{Procedia computer science},
  vol. 147, pp. 400--406, 2019.

\bibitem{zhou2018stock}
X.~Zhou, Z.~Pan, G.~Hu, S.~Tang, and C.~Zhao, ``Stock market prediction on
  high-frequency data using generative adversarial nets,'' \emph{Mathematical
  Problems in Engineering}, vol. 2018, 2018.

\bibitem{takahashi2019modeling}
S.~Takahashi, Y.~Chen, and K.~Tanaka-Ishii, ``Modeling financial time-series
  with generative adversarial networks,'' \emph{Physica A: Statistical
  Mechanics and its Applications}, vol. 527, p. 121261, 2019.

\bibitem{gould2013limit}
M.~D. Gould, M.~A. Porter, S.~Williams, M.~McDonald, D.~J. Fenn, and S.~D.
  Howison, ``Limit order books,'' \emph{Quantitative Finance}, vol.~13, no.~11,
  pp. 1709--1742, 2013.

\bibitem{graves2013generating}
A.~Graves, ``Generating sequences with recurrent neural networks,'' \emph{arXiv
  preprint arXiv:1308.0850}, 2013.

\bibitem{bengio2015scheduled}
S.~Bengio, O.~Vinyals, N.~Jaitly, and N.~Shazeer, ``Scheduled sampling for
  sequence prediction with recurrent neural networks,'' \emph{arXiv preprint
  arXiv:1506.03099}, 2015.

\bibitem{goodfellow2014generative}
I.~J. Goodfellow, J.~Pouget-Abadie, M.~Mirza, B.~Xu, D.~Warde-Farley, S.~Ozair,
  A.~Courville, and Y.~Bengio, ``Generative adversarial networks,'' \emph{arXiv
  preprint arXiv:1406.2661}, 2014.

\bibitem{cont2014price}
R.~Cont, A.~Kukanov, and S.~Stoikov, ``The price impact of order book events,''
  \emph{Journal of financial econometrics}, vol.~12, no.~1, pp. 47--88, 2014.

\bibitem{hochreiter1997long}
S.~Hochreiter and J.~Schmidhuber, ``Long short-term memory,'' \emph{Neural
  computation}, vol.~9, no.~8, pp. 1735--1780, 1997.

\bibitem{zhang2015sensitivity}
Y.~Zhang and B.~Wallace, ``A sensitivity analysis of (and practitioners' guide
  to) convolutional neural networks for sentence classification,'' \emph{arXiv
  preprint arXiv:1510.03820}, 2015.

\bibitem{dunnett1992step}
C.~W. Dunnett and A.~C. Tamhane, ``A step-up multiple test procedure,''
  \emph{Journal of the American Statistical Association}, vol.~87, no. 417, pp.
  162--170, 1992.

\bibitem{jarque1980efficient}
C.~M. Jarque and A.~K. Bera, ``Efficient tests for normality, homoscedasticity
  and serial independence of regression residuals,'' \emph{Economics letters},
  vol.~6, no.~3, pp. 255--259, 1980.

\end{thebibliography}

\end{document}